\documentclass[preprint,5p]{elsarticle}

\usepackage{hyperref}
\usepackage{lineno}
\modulolinenumbers[5]

\journal{Energy and Buildings}

\usepackage{eurosym}
\usepackage{amssymb}
\usepackage{amsmath}
\usepackage{graphicx}
\usepackage{tikz}
\usetikzlibrary{calc}
\usepackage[siunitx, american]{circuitikz}
\usepackage{subfig}
\usepackage{array}
\usepackage{algorithm}
\usepackage{algorithmic}

\begin{document}

\begin{frontmatter}

\title{The use of distributed thermal storage in district heating grids for demand side management}


\author[VITO,EVille]{D.~Vanhoudt\corref{cor1}}
\author[VITO,EVille,REstore]{B.J.~Claessens}
\author[VITO,EVille]{R.~Salenbien}
\author[VITO,EVille]{J.~Desmedt}
\cortext[cor1]{Corresponding author: tel.: +32-14-33-59-74; e-mail address: dirk.vanhoudt@vito.be (D. Vanhoudt)}
\address[VITO]{Flemish Institute for Technological Research (VITO), Boeretang 200, B-2400 Mol, Belgium}
\address[EVille]{EnergyVille, Thor Park 8310, B-3600 Genk, Belgium}
\address[REstore]{REstore, Posthofbrug 12, B-2600 Antwerp, Belgium}

\begin{abstract}
The work presented in this paper relates to a small scale district heating network heated by a gas fired CHP. In most common situations, such a CHP is heat driven operated, meaning that the CHP will switch on whenever heat is needed, while not taking into account the demand of electricity at that time. In this paper however, an active control strategy is developed, aiming to maximize the profit of the CHP, selling its electricity to the spot market. The CHP will therefore switch on at moments of high electricity prices. Nevertheless, since there never is a perfect match between the demand of heat and the demand of electricity, a thermal energy storage is included in the network to overcome the difference between supply and demand of heat in the network. In this study, three different storage concepts are compared: (1) a central buffer tank next to the CHP; (2) small storage vessels distributed over the different connected buildings; and (3) the use of the thermal mass if the buildings as storage capacity. 
Besides the development of the control algorithms based on model predictive control, a simulation model of the network is described to evaluate the performance of the different storage concept during a representative winter week. The results show that the presented control algorithm can significantly influence the heat demand profile of the connected buildings. As a results, active control of the CHP can drastically increase the profit of the CHP. The concept with the distributed buffers gives the best results, however the profit for the thermal mass concept is only marginally smaller. Since in this latter case no significant investment costs are needed, the conclusion for this case study is that the use of thermal mass of buildings for demand side management in district heating systems is very promising.
\end{abstract}


\end{frontmatter}


\section{INTRODUCTION}
\subsection{Background}
District heating networks (DHN) are collective systems used for heating of buildings. The infrastructure typically consists of one or more heat production plants, transport and distribution pipes, and substations delivering the heat to the customers. A DHN is an alternative for the individual heat production by means of a gas boiler for example. District heating is widely spread especially in northern, central and eastern Europe. E.g. in the Scandinavian and Baltic countries the share of district heating in the heat demand is close to or above 50\% \cite{Euroheat&Power2016}.

Compared to individual heat production district heating has some benefits. District heating systems allow valorisation of surplus heat from industrial processes or waste incineration and therefore increase the energy efficiency \cite{Fang2013,Holmgren2006a}. It also facilitates the transition to renewable heating of buildings, since renewable sources of heat (biomass, geothermal energy, solar thermal energy...) are often large-scale or at least too expensive for small scale applications as single building heating \cite{Lund2009b}. Therefore also in other countries, DHNs are perceived to play an increasingly important role in the future energy infrastructure \cite{Munster2012, Lund2010}.

Also the generation of electricity is getting more sustainable. Renewable energy sources (RES) have a growing share in the total electricity production. In 2011 this share was 27.5\% in the EU-28 \cite{REN212016}. By 2050, the share of renewables should further increase to 48\% according the IEA Blue Map scenario \cite{InternationalEnergyAgency2010}. An unfavourable aspect of intermittent renewable sources like wind and photovoltaic however is that they are highly fluctuating and therefore partly unpredictable and uncontrollable. This aspect of these RES also effects the spot price of electricity on the energy markets. Studies indicate a decrease of the average spot price \cite{Sensfuß2008,Woo2011b}, but also an increase of the volatility of the price \cite{Green2010a,Milstein2011a}.

In today's European district heating grids, about three quarter of the total heat supply is supplied by combined heat and power plants (CHP) \cite{DHC+TechnologyPlatform2009a}. The electricity produced simultaneously with the heat production can be sold on the spot markets. Since the increased variance in the price of electricity, for the profitability of the CHP plant it is important to produce electrical power when the price is high. In this way, the CHP indirectly contributes to the balance of the electricity grid: when a lot of intermittent renewable power is available the remaining power demand and correspondingly the price will be low, stimulating the CHPs to switch off. In the same way, scarcity of renewable energy will invoke the CHPs to switch on.

Off course, the demand of thermal power will never be fully synchronised to the demand of electricity. Therefore, thermal energy storage is required when controlling the CHP electricity driven. Usually, this storage consists of one large water storage tank placed next to the CHP.

In this paper, a case study is presented of a small scale district heating system heated by a CHP. The CHP is controlled 'actively'. This means that, instead of regular heat driven control of the CHP, the control algorithm tries to meet a certain objective, in this case to maximise the profit for the CHP owner. In practice, this means that the CHP will be enabled as much as possible at times of high electricity price. The heat produced simultaneously is stored in a thermal energy storage. Besides the usual control storage concept described above, also a configuration with distributed storage tanks is taken into account. Finally, a configuration whereby the thermal mass of buildings is used to store thermal energy. The research question in this paper is straightforward: which storage concept performs best for the studied case?

\subsection{Operational optimization of district heating CHPs}
\label{sec:lit_optimization} Research on active control of district heating CHPs is far from new. The last decade, several studies were published on CHPs selling their electricity to the spot market. In \cite{Brujic2007} the simulation results are shown for the optimal control of a CHP with a thermal storage. For different situations the cost savings are calculated. Compared to the case whereby the CHP is following the heat demand, the cost saving due to active control amounts to 17\% for example. Also Rolfsman \cite{Rolfsman2004, Cho2009} presented an optimization algorithm to maximize the profit of a CHP based on mixed integer linear programming. Mixed integer linear programming was already used in 1992, when Gustafsson and Karlsson \cite{Gustafsson1992} developed an optimization scheme for the operation of a district heating CHP. This scheme was applied to a case study in Malm\"{o}, Sweden. Also Gunkel et al. \cite{Gunkel2012} applied a mixed integer linear programming technique to schedule the operation of a fleet of small CHPs in a micro-grid. Others used a combination of dynamic programming and Lagrangian relaxation techniques \cite{Dotzauer1997}, fuzzy linear programming \cite{Lee1999}, or heuristic optimization methods \cite{Perea2016}.

Some authors also employ existing optimization solvers to schedule CHP operation. E.g in \cite{Schaumburg-Muller2006} GAMS is used to solve the scheduling problem of a CHP and \cite{Bogdan2006} describes the use of ACOM. Another very popular scheduling software is energyPRO \cite{EMD}, used in studies like e.g. \cite{Andersen2007a, Streckiene2009}.

\subsection{Goal of the tests}
\label{sec:goals_of_tests}
All studies mentioned in \ref{sec:lit_optimization} assume a central buffer, placed at the CHP plant. This is the most obvious solution, but other options can be thought of as well. In this work, we quantify the profit of switching district heating CHPs on appropriate moments (high spot market price) and compare the performance for different storage possibilities. Therefore a simulation model was developed for a fictive DHN in a neighbourhood with 100 connected buildings. A representative winter week was simulated for a reference case and the different storage cases. In the reference case, the CHP is heat-driven controlled. No thermal storage is present and the CHP always produces the actual heat demand of the heat district grid. In this case, electricity is a surplus product sold at any price. Besides this reference case, the three storage configurations are evaluated:

\begin{enumerate}
	\item In this configuration, next to the CHP a central storage tank is installed. The CHP is electricity driven, meaning that the operation of the CHP is optimized to the actual stop price of electricity. The big advantage of this configuration is that there is no intervention in the control of the DHN valves in the building substations. When the building needs heat at a certain time, the valve of the district heating opens until no heat is needed any more. Also, no communication is needed between the CHP control software and the individual buildings.
	\item In the second configuration, the thermal storage capacity is distributed amongst the individual buildings, by installing small buffers in every building. When it is interesting to switch on the CHP, the control system will open the valves of the building substations with the highest actual heat demand, i.e. the building with the most empty buffer. Since in this configuration communication is needed between the individual buildings and central control system anyway, also the thermal mass of the building can be activated. This is established by increasing the indoor temperature of the building within certain limits. In this configuration, the total volume of all buffer tanks together is the same as the central buffer in the first storage configuration but the additional thermal mass of the building increases the storage capacity compared to the first configuration. 
	\item The last configuration is included to quantify the potential of the activation of building mass only. This configuration is similar to the second one, only the distributed buffers are omitted. 
\end{enumerate}

The first storage configuration is most common and was applied in the above mentioned studies \cite{Brujic2007,Rolfsman2004,Gunkel2012,Dotzauer1997,Lee1999,Schaumburg-Muller2006,Bogdan2006,Andersen2007a, Streckiene2009}. Storage in the thermal mass in buildings (the third configuration) was studied in \cite{Ingvarson2008,Johansson,Rolfsman2004,Reynders2014}. Distributed storage vessels as in the second configuration are a lot less common in literature. They are mentioned though as a solution to deploy district heating in areas with low heat densities \cite{Zinko2008}.

\section{METHODS}
\subsection{The district heating pipe model}
\label{sec:network_model}
To calculate the flow rates and pressures in the network, a method developed by Valdimarsson \cite{Valdimarsson1993} is used. In analogy to electrical circuits the Kirchoff laws are applied, whereby voltage is replaced by pressure and flow rate by current. A complication is however that in hydraulic circuits the relation between flow rate and pressure is not linear, like the relation between voltage and current in electrical circuits (Ohm’s law), but quadratic. \par
To calculate the temperature evolution in a pipe, the node model developed by Benonysson \cite{Benonysson1991} was implemented. This quasi-dynamic model relies on the fact that pressure and flow in district heating grids change orders of magnitudes faster than the temperature of the water in the pipes. In this model, the outlet temperature of a pipe is calculated as the input temperature of a certain amount of time steps before. Then, a correction is applied to this calculated outlet temperature, to take into account the heat capacity of the pipe wall and heat losses to the environment. A summary of this method can be found in \cite{Palsson1999}.

\begin{figure}[h]
	\centering
	\includegraphics[width=.4\textwidth]{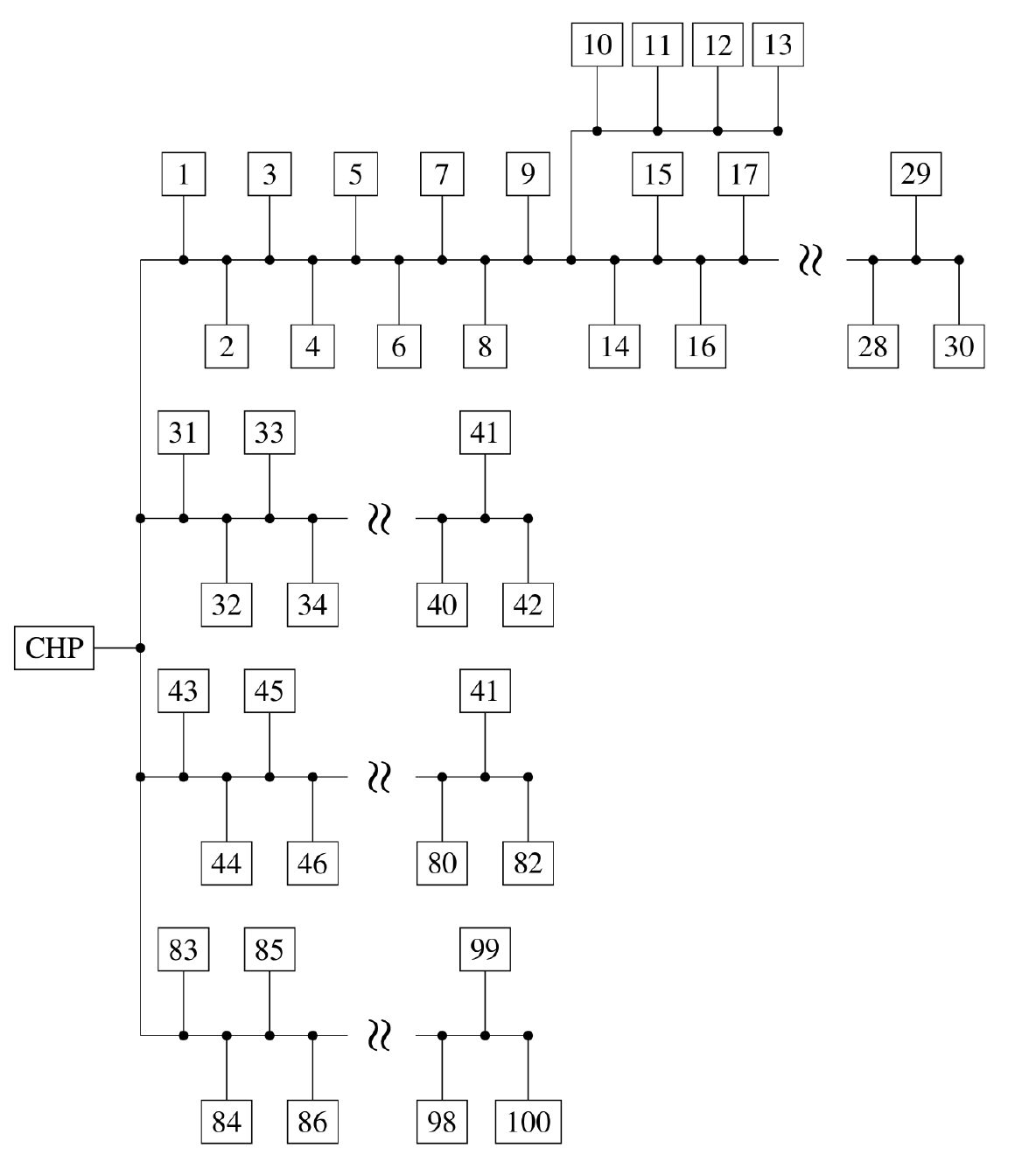}
	\caption{Schematic representation of the district heating grid}
	\label{fig:network}
\end{figure}

An schematic representation of the district heating network studied in this paper is shown in figure \ref{fig:network}. The CHP provides heat to 100 buildings, located in 4 streets. The network consists of twin pipes with a total length of 2.1 km and pipe diameters ranging from DN25 to DN100. With these diameters, the pressure losses in the pipes are limited to $200 \ Pa/m$.

\subsection{The building model}
\begin{figure*}[h]
	\centering
	\includegraphics[width=.7\textwidth]{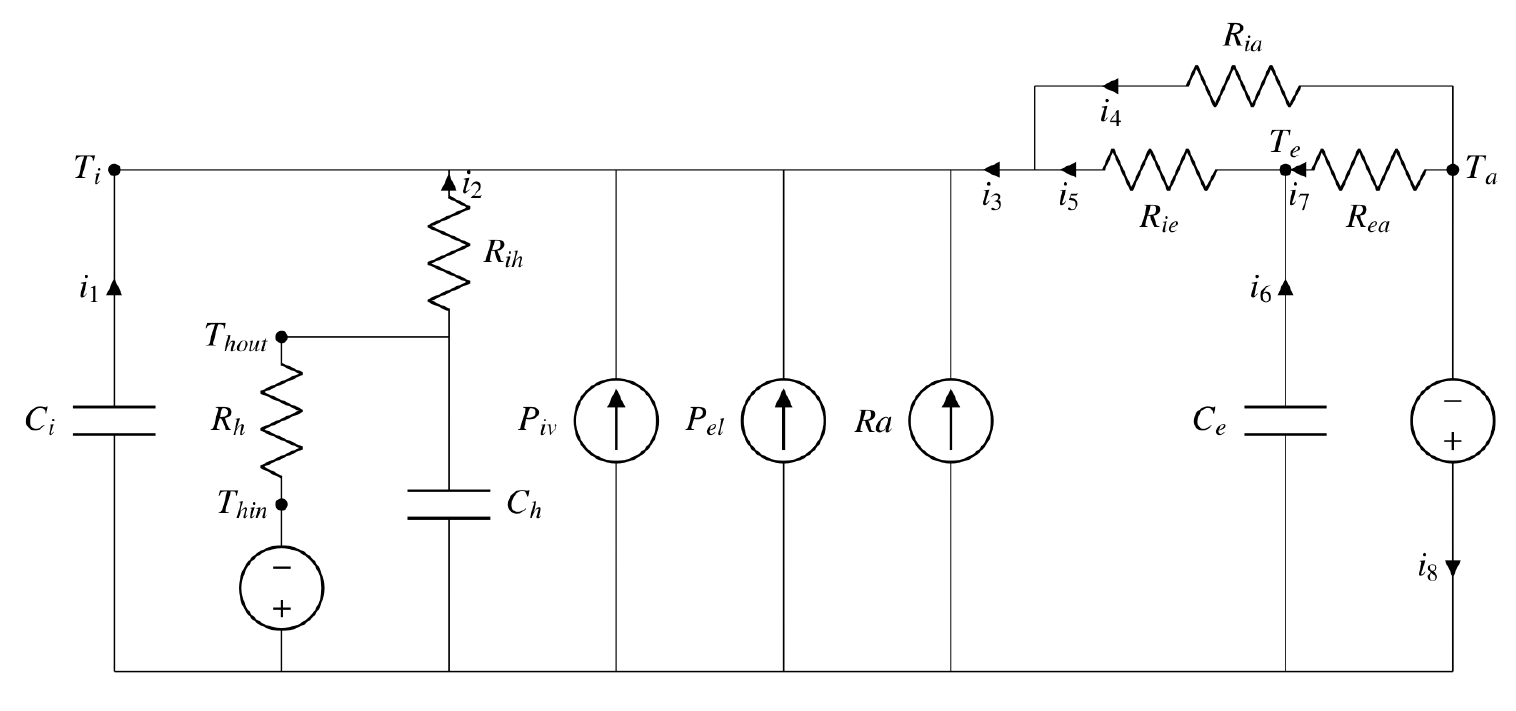}
	\caption{RC-circuit of the building model}
	\label{fig:RC}
\end{figure*}

Every building in the network is represented by a lumped capacitance model. In these models the thermal problem is translated to an electric analogue, whereby a temperature is transformed to a voltage and power to an electrical current. The building properties are described as a combination of resistances (R) and capacitances (C). By solving the Kirchoff equations, the evolution of the temperatures in the buildings are calculated. The buildings in the studied network all have the same circuit, shown in figure \ref{fig:RC}, but the values of the parameters differ for every house.

Applying the Kirchoff current law leads to equations \ref{eq:Kirchoff}. By solving these equations, the evolution of the temperatures can be calculated.
\begin{equation}
\left\{
\begin{array}{l}
-C_i \frac{dT_i}{dt} + \frac{T_{hout} - T_i}{R_{ih}} + P_{iv} + P_{el} + Ra + \frac{T_a - T_i}{R_{ia}} + \frac{T_e - T_i}{R_{ie}} = 0 \\[5mm]
-C_e \frac{dT_e}{dt} + \frac{T_a - T_e}{R_{ea}} = \frac{T_e - T_i}{R_{ie}} \\[5mm]
-C_h \frac{dT_{hout}}{dt} + H\frac{T_{hin} - T_{hout}}{R_h} = \frac{T_{hout} - T_i}{R_{ih}} 
\end{array}
\right.
\label{eq:Kirchoff}
\end{equation}

In table \ref{tab:RCparams} an overview is given of the meaning of the different parameters, inputs and outputs of the model as well as the units used in the simulations. The remains if this section discusses the determination of the values of the parameters.

\begin{table}[h!]
\begin{center}
\begin{tabular}{l l l}
\hline
symbol 			& description								& unit  			\\
\hline
$R_h$			& thermal resistance radiator				& $^\circ C/kW$		\\
$R_{ih}$		& thermal resist. radiator-inside		& $^\circ C/kW$ 	\\
$R_{ie}$		& thermal resist. inside-envelope		& $^\circ C/kW$ 	\\
$R_{ea}$		& thermal resist. envelope-ambient	& $^\circ C/kW$ 	\\
$R_{ia}$		& thermal resist. inside-ambient		& $^\circ C/kW$ 	\\
$C_i$			& thermal capacitance inside				& $kWh/^\circ C$ 	\\
$C_h$			& thermal capacitance radiator				& $kWh/^\circ C$ 	\\
$C_e$			& thermal capacitance envelope				& $kWh/^\circ C$ 	\\
$H$				& boolean: 1 if heating is on		& -					\\
$T_a$			& ambient temperature						& $^\circ C$		\\
$P_{iv}$		& power infiltration						& $kW$				\\
$P_{el}$		& electrical power building					& $kW$				\\
$Ra$			& power of solar irradiation				& $kW$				\\
$T_i$			& building inside temperature				& $^\circ C$		\\
$T_e$			& building envelope temperature				& $^\circ C$ 		\\ 
$T_{hout}$		& heating system return temp.			& $^\circ C$		\\
\hline 
\end{tabular}
\caption{parameters, inputs and outputs of the RC-circuit of the building}
\label{tab:RCparams}
\end{center}
\end{table}

The 100 buildings used in the simulation are all derived from one standard building, in detail described in \cite{Verbeek2007}. This building is a detached house with a living area of 103 $m^{2}$ and a protected volume of 452 $m^{3}$. The building has a K-value of 40, which is the legal norm in Flanders. For this building $C_i$ is 20.13 $kWh/K$,  and $C_e$ is 21.23 $kWh/K$. The building is heated by means of radiators which have a low thermal mass, resulting in a thermal capacitance $C_h$ of 0.17 $kWh/K$.

Air infiltration losses are modelled by as a power $P_{iv}$ and resistance $R_{ia}$. The power loss $P$ due to infiltration can be calculated by the formula:
\begin{eqnarray} 
P &=& -\dot{m} \ c_p \ \Delta t   \\
  &=& - \rho \dot{V} \ c_p \Delta t
\end{eqnarray}
where $\dot{m}$ is the mass flow rate of the air flow due to infiltration, $\dot{V}$ the volume flow rate, $c_p$ and $\rho$ the heat capacity and density of air and $\Delta t$ the temperature difference between inside and outside.
$\dot{V}$ was determined by the formula \cite{AmericanSocietyofHeatingRefrigerationandAir-ConditioningEngineers2009}:
\begin{equation}
\dot{V} = A_L \sqrt{C_S \Delta t + C_W U^2}
\end{equation}
where $A_L$ is the affective air leakage area, $C_S$ is the stack coefficient and $C_W$ the wind coefficient and $U$ the wind speed. Linearisation of this formula leads to:
\begin{equation}
P = \frac{\Delta T}{R_{ia}} + A_{Piv}.U + B_{Piv}
\end{equation}
with
\begin{equation}
\begin{array}{l}
\displaystyle R_{ia} = - \Bigg[A_L \ \rho \ c_p \ \frac{C_S+2 C_S \Delta t_0 + 2 C_W U_0^2}{2 \sqrt{C_S \Delta t_0 + C_W U_0^2}} \Bigg]^{-1} \\[7mm]
\displaystyle A_{Piv} = -A_L \ \rho \ c_p \ \frac{C_w \ U_0 \ \Delta T_0}{\sqrt{C_S \Delta t_0 + C_W U_0^2}} \\[7mm]
\displaystyle B_{Piv} = A_L \ \rho \ c_p \frac{C_s + 2C_w \ U_0^2 \ \Delta T_0}{2\sqrt{C_S \Delta t_0 + C_W U_0^2}}
\end{array}
\end{equation}
Finally $P_{iv}$ is defined as $P_{iv} \equiv A_{Piv}.U + B_{Piv}$. In the simulation, the following values are used for the coefficients: $C_S = 4.35 \ 10^{-4}$, $C_W = 1.61 \ 10^{-4}$, $A_L = 621 \ 10^{-4} m²$, $\Delta T_0 = 12.5^\circ C$ and $U_0 = 3.5 \ m/s$.

Calculated according EN 12831 \cite{EuropeanCommitteforStandardisation2003}, the reference building has maximum static power demand of 9.8 kW at an internal temperature of $20^\circ C$ and an ambient temperature of $-8^\circ C$, and excluding ventilation and infiltration losses ($R_{ia}=0$), electrical power ($P_{el}=0$) and solar irradiation ($Ra=0$). From the scheme in \ref{fig:RC}, one then can deduct $R_{ie} + R_{ea} = (20^\circ C - (-8)^\circ C)/(9.8 kW) = 2.87 ^\circ C/kW$. For the reference building $R_{ie} = 1^\circ C/kW$. 

Once all the values for the coefficients are known, the design power for the heating system can be determined. Therefore to the maximum static power demand of 9.8 kW the infiltration losses are added, as well as a reheat term to allow fast reheat of the building. This reheat term was chosen at $22 W/m^2$. In this way, the heating system design load for the standard building is 16.0 kW. The design temperature regime for the heating system is $70^\circ C-30^\circ C$, therefore $R_h$ can be calculated as $R_h = (70-30)/16.0 ^\circ C/kW = 2.49 ^\circ C/kW$. In the same way, $R_{ih}$ can be defined as $R_{ih} = (30-20)/16.0 ^\circ C/kW = 0.62 ^\circ C/kW$.

The values above are determined for the standard building. Since in the simulation 100 buildings are used it is relevant to introduce statistical spread in the building parameters. Therefore, for the different buildings the parameters $C_i$, $C_e$, $C_h$, $C_W$, $C_S$, $R_{ie}$ and $R_{ea}$ are normally distributed around the value of the standard building with a standard deviation of 20\% of the mean value. An overview of the resulting values used in the simulations is shown in \ref{tab:RCvalues}.

\begin{table}[h!]
\begin{center}
\begin{tabular}{l r r r}
\hline
parameter 	& mean		& min		& max	  	\\
\hline
$R_h$		& 2.4825 	& 1.5881 	& 3.6914	\\
$R_{ih}$	& 0.6206 	& 0.3970 	& 0.9229	\\
$R_{ie}$	& 1.0035 	& 0.5332 	& 1.7233	\\
$R_{ea}$	& 2.0057 	& 0.9350 	& 5.0462 	\\
$R_{ia}$	& 12.3195 	& 4.6360 	& 25.5365	\\
$C_i$		& 20.2141 	& 10.0500 	& 30.9041	\\
$C_h$		& 0.1621 	& 0.0936  	& 0.2301	\\
$C_e$		& 21.2738 	& 12.4910 	& 30.5981	\\
\hline 
\end{tabular}
\caption{values of the RC-circuit parameters of the building}
\label{tab:RCvalues}
\end{center}
\end{table}

\subsection{The central buffer model}
All buffers in the simulations are modelled by means of a multi-node model. This is a one-dimensional model whereby the buffer vessel is represented as number of stacked volume segments. Each segment is fully-mixed, meaning that the whole segment is assumed to have the same temperature. A mathematical description of the model is described in \cite{Kleinbach1993, Newton1995}. In the simulation described in this paper, the central buffer has 50 layers.

\begin{figure}[h]
	\centering
	\includegraphics[width=.4\textwidth]{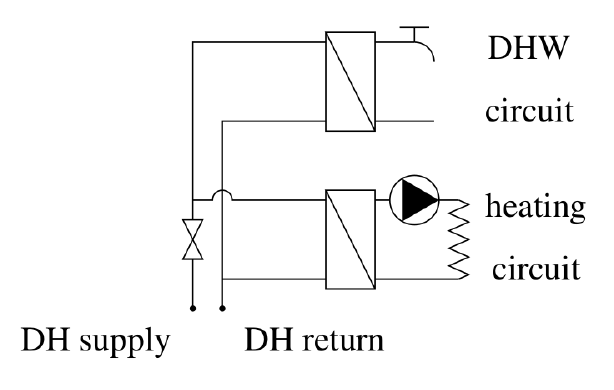}
	\caption{substations in configurations without local buffers}
	\label{fig:substations}
\end{figure}

In the configuration with a central storage buffer, there are no storage vessels installed in the houses. In this case, the coupling between the building and the district heating network is achieved by means of a commonly used indirect substation set as shown in figure \ref{fig:substations}.

\subsection{The distributed buffers model}
In the configuration with the distributed buffers, three types of buffers are considered, randomly distributed amongst the buildings. These types are representative for decentralised buffers on the market today.

\begin{figure*}[h]
	\centering
	\includegraphics[width=.9\textwidth]{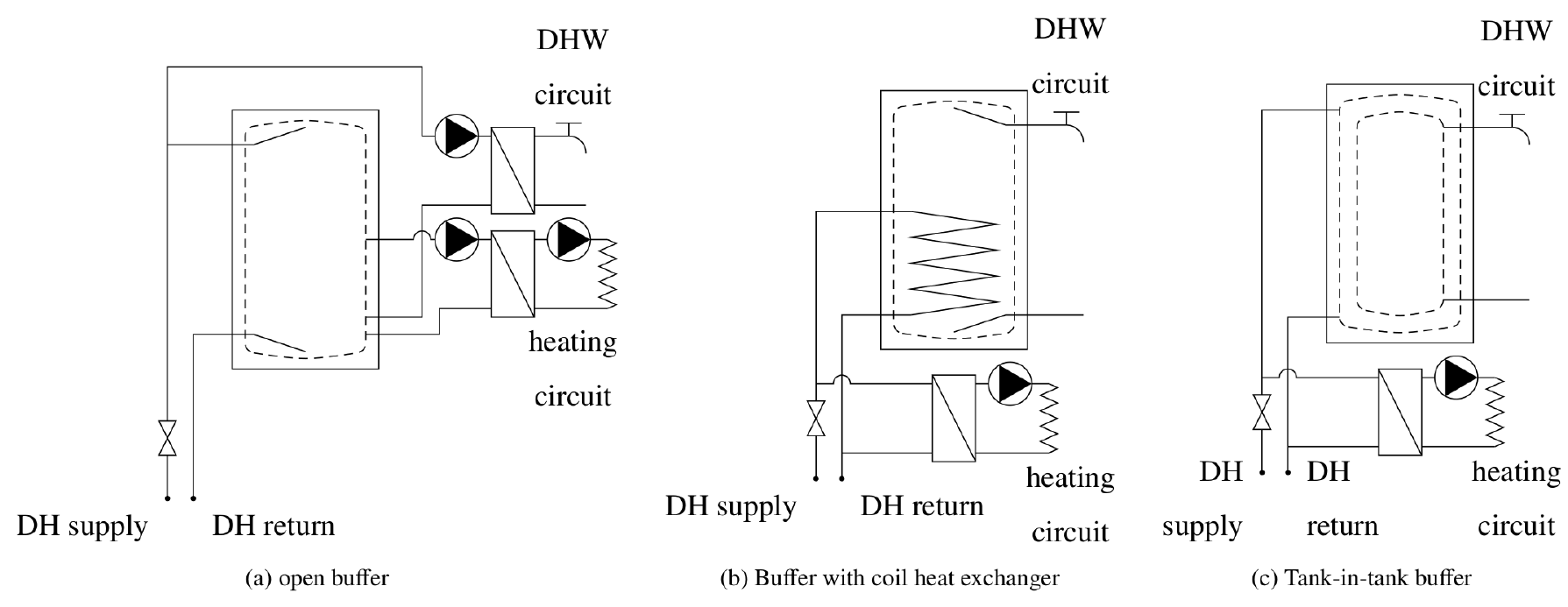}
	\caption{Buffer types used in the simulation}
	\label{fig:buffer_types}
\end{figure*}

The first type is an open buffer type, as shown in figure \ref{fig:buffer_types}a. This is the most simple type of buffer, with no separation between the primary and secondary flow of the buffer. In this buffer, heat is stored for building heating as well as domestic hot water production. Two heat exchangers are provided to separate the district heating water from the building's heating circuit and the domestic hot water. The buffers in the simulation have a volume of 500 l. 

The second type is a buffer with an immersed coil heat exchanger, as presented in figure \ref{fig:buffer_types}b. This type of buffer is commonly used in solar thermal installations, where the coil heat exchanger realizes a physical barrier between the heat transfer fluid and the domestic hot water. A buffer volume of 200 l is used in the simulations. 

The last type of buffer as a tank-in-tank buffer (figure \ref{fig:buffer_types}c), whereby the water from inner tank is heated by the fluid in the other tank. The tank wall between the inner and outer tank acts as a heat exchanger in this case. In the model, the inner tank has a volume 164 l while the outer tank's volume is 39 l. In these cases, heat for space heating is not stored, so the demand of heat needs to be fulfilled directly by the district heating grid. 

Like in the configuration with a central buffer vessel, they are modelled by means of  a multi-node model. To limit the computational time, the number of layers per buffer is chosen lower than for the centralised buffer: in this configuration, each local buffer has 15 layers.

\subsubsection{The CHP and boiler model}
The heat to the network is provided by a natural gas-fired CHP, with a natural gas boiler as backup. These components are represented by means of quasi-static black-box models. The equations for the CHP are:

\begin{equation}
\left\{
\begin{array}{l}
\displaystyle P_{el} = P_{el \ max} \ f_m, \quad f_{m \ min} \leqslant f_m \leqslant 1\\[5mm]
\displaystyle P_{heat} = A_{heat}(T_{in}) \ f_m + B_{heat}(T_{in}) \\[5mm]
\displaystyle P_{gas} = \alpha_{gas} \ f_m^2 + \beta_{gas} \ f_m + \gamma_{gas} \\[5mm]
\displaystyle A_{heat}(T_{in}) = \alpha_{heat} \ T_{in}^3 + \beta_{heat} \ T_{in}^2 + \gamma_{heat} \ T_{in} + \delta_{heat} \\[5mm]
\displaystyle B_{heat}(T_{in}) = \epsilon_{heat} \ T_{in}^2 + \zeta_{heat} \ T_{in} + \eta_{heat}
\end{array}
\right.
\label{eq:CHP}
\end{equation}
where $P_{el}$ and $P_{heat}$ are the electrical and thermal power produced by the CHP, $P_{gas}$ is the input power consumption. $P_{el \ max}$ is the maximum electrical power of the CHP (parameter), $f_m$ the modulation factor of the CHP which is limited between $f_{m \ min}$, the lower modulation limit and $1$. $\alpha_{gas}$ to $\gamma_{gas}$ and $\alpha_{heat}$ to $\eta_{heat}$ are parameters, fitted to partial load curves supplied by a CHP-supplier. Apart from the calculation of these equations, a minimum on- and off-time ($\Delta t_{on \ min}$ and $\Delta t_{off \ min}$) is added to the model. This means that, once the CHP is active, the CHP must stay on during a certain time. In the same way, once the CHP shuts down it stays off during a certain period. The parameters used in the simulation are shown in table \ref{tab:CHP_params}.

\begin{table}[h!]
\begin{center}
\begin{tabular}{l r l}
\hline
parameter 					& value 				& unit	\\
\hline
$P_{el \ max}$				& $600$ 				& $kW$	\\
$f_{m \ min}$				& $0.4$ 				& $-$	\\
$\Delta t_{off \ min}$ 		& $15$ 					& $min$ \\
$\Delta t_{on \ min}$ 		& $15$ 					& $min$ \\
$\alpha_{gas}$				& $31.250$ 				& $kW$	\\
$\beta_{gas}$				& $1310.75$ 			& $kW$ 	\\
$\gamma_{gas}$				& $181.35$ 				& $kW$	\\
$\alpha_{heat}$				& $3.1537 \ 10^{-5}$ 	& $kW/(^\circ C)^3$	\\
$\beta_{heat}$				& $-7.4162 \ 10^{-3}$	& $kW/(^\circ C)^2$	\\
$\gamma_{heat}$				& $-0.3258$ 			& $kW/^\circ C$	\\
$\delta_{heat}$				& $704.09$ 				& $kW$	\\
$\epsilon_{heat}$			& $6.0633 \ 10^{-4}$ 	& $kW/(^\circ C)^2$	\\
$\zeta_{heat}$				& $-0.1848$ 			& $kW/^\circ C$	\\
$\eta_{heat}$				& $160.01$ 				& $kW$	\\
\hline 
\end{tabular}
\caption{values of the CHP parameters}
\label{tab:CHP_params}
\end{center}
\end{table}

Similarly, also for the gas boiler a quasi-static black-box model was developed. 

\begin{equation}
\left\{
\begin{array}{ll}
\displaystyle P_{gas} = & P_{gas \ nom} \ f_m, \quad f_{m \ min} \leqslant f_m  \\[5mm]
\displaystyle P_{out} =  &eff \ . \ P_{gas} \\[5mm]
\displaystyle eff = & A(P_{gas})*T_{in}^3 + B(P_{gas})*T_{in}^2 \\
& + C(P_{gas})*T_{in} + D(P_{gas}) \\[5mm]
\displaystyle A(P_{gas}) = & \alpha \ P_{gas}^2 + \beta \ P_{gas} + \gamma \\[5mm]
\displaystyle B(P_{gas}) = & \delta \ P_{gas}^2 + \epsilon \ P_{gas}+ \zeta \\[5mm]
\displaystyle C(P_{gas}) = & \eta \ P_{gas}^2 + \theta \ P_{gas} + \iota  \\[5mm]
\displaystyle D(P_{gas}) = & \kappa \ P_{gas}^2 + \mu \ P_{gas} + \nu 
\end{array}
\right.
\label{eq:boiler}
\end{equation}
where $P_{heat}$ is the heat output, $P_{gas \ nom}$ the nominal gas input, $P_{gas}$ the gas input $\alpha$ to $\nu$ are parameters again, fitted to supplier data. $f_{m \ min}$ is the lower modulation limit. In table \ref{tab:boiler_params} an overview is given of the parameters used.

\begin{table}[h!]
\begin{center}
\begin{tabular}{l r l}
\hline
parameter 				& value 				& unit	\\
\hline
$P_{gas \ nom}$			& $1100$ 				& $kW$	\\
$f_{m \ min}$			& $0.1$ 				& $-$	\\
$\alpha$				& $-7.758 \ 10^{-13}$ 	& $(^\circ C^3 \ kW^2)^{-1}$	\\
$\beta$					& $ -1.119 \ 10^{-10}$	& $(^\circ C^3 \ kW)^{-1}$	\\
$\gamma$				& $3.295 \ 10^{-6}$ 	& $(^\circ C^3)^{-1}$	\\
$\delta$				& $1.195 \ 10^{-10}$ 	& $(^\circ C^2 \ kW^2)^{-1}$	\\
$\epsilon$				& $2.911 \ 10^{-8}$ 	& $(^\circ C^2 \ kW)^{-1}$	\\
$\zeta$					& $-4.665 \ 10^{-4}$ 	& $(^\circ C^2)^{-1}$	\\
$\eta$					& $-6.067 \ 10^{-9}$ 	& $(^\circ C \ kW^2)^{-1}$	\\
$\theta$				& $-1.558 \ 10^{-6}$ 	& $(^\circ C \ kW)^{-1}$	\\
$\iota$					& $1.800 \ 10^{-2}$ 	& $(^\circ C )^{-1}$	\\
$\kappa$				& $1.121 \ 10^{-7}$ 	& $(kW^2)^{-1}$	\\
$\mu$					& $ -1.503 \ 10^{-5}$ 	& $(kW)^{-1}$	\\
$\nu$					& $7.675 \ 10^{-1}$ 	& $-$	\\
\hline 
\end{tabular}
\caption{values of the gas boiler parameters}
\label{tab:boiler_params}
\end{center}
\end{table}

\subsection{The control algorithms}
\subsubsection{Reference case}
\label{sec:reference_control}
\paragraph{The local building control system} ~\\
In the reference case no storage is applied. All buildings use a common thermostat control system to maintain the indoor temperature settings. This means that, when the building indoor temperature drops below a set point, the valve from the district heating grid is opened and heat is supplied to the building until the indoor temperature raises above another set point. The heating circuit supply set temperature is determined by means of a heating curve. This set point is maintained by adjusting the flow rate of the district heating water with the district heating valve. When there is a domestic hot water demand, the district heating valve is fully opened. When both a heating demand and domestic hot water demand exists, the district heating water flow rate is divided over the two circuits.

\paragraph{The CHP and gas boiler control system} ~\\
The set temperature of the district heating grid supply temperature is determined by a heating curve, using the mean outside temperature during the past 24 hours as an input. The CHP is than modulated to this set temperature. If the set temperature cannot be reach by the CHP, the gas boiler is enabled as well. Also if the desired power of the CHP is below the lower modulation limit of the installation, the CHP shuts down and the gas boiler is used to supply the necessary heat.

\subsubsection{Active control cases}
\label{sec:active_control_algorithm}

\begin{figure}[h]
	\centering
	\includegraphics[width=.4\textwidth]{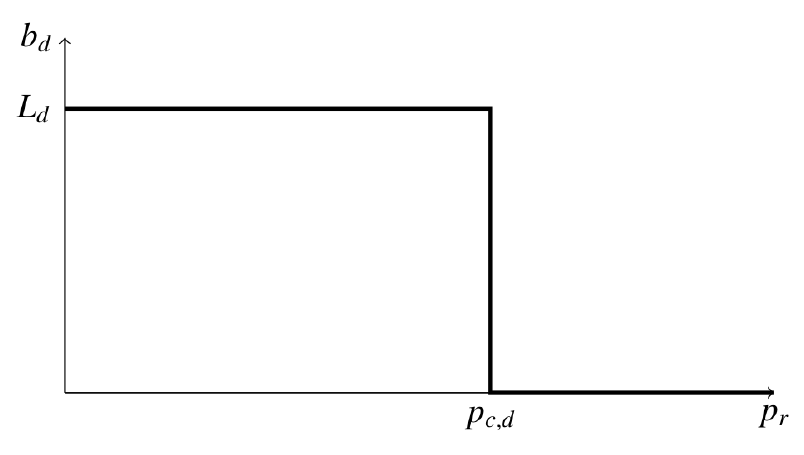}
	\caption{The bidding curve of the TLCs.}
	\label{fig:bidding_curve}  
\end{figure}

The active control algorithm is based on a \textit{pragmatic} Model Predictive Control (MPC) approach \cite{Camacho2004} following the Three Step Approach (TSA) as presented in \cite{Stijn,Arnout}. The water buffers and the thermal mass of the buildings in this work, can be seen as Thermostatically Controlled Loads (TCLs) as in \cite{ClaessensSelfLearning}. Here however, an extension must be made for TCLs coupled to a DHN, as explained in \cite{Claessens2017}. 
In the first aggregation step, the temperature information of the TCLs is collected, as is a bid function as explained below. Then, these individual bid functions are aggregated (summarised). In the second step, a control action is generated based upon an optimization spanning a finite time horizon $T$. In the third step, this control action is projected on local decisions using a market-based multi-agent system. For the first scenario with a central hot water storage tank the TLCs at building level follow the default control strategy as detailed in Section \ref{sec:reference_control}. 
These three steps are repeated every time step $t$.
\paragraph{Step 1: Aggregation} ~\\
\label{s.stepOne}
At every time step $t$, bid functions $b_{d}(p_{r})$ are constructed, whereby $p_{r}$ represents the priority for the TCL, which could be seen as a 'virtual price' which the TCL is willing to pay for heat (Figure \ref{fig:bidding_curve}).

To do so, first  the temperature $T_{d,t}$ information is retrieved from all storage systems $d \in \mathcal{S}$. For the water storage buffers, in configurations 1 and 2 (section \ref{sec:goals_of_tests}), $T_{d,t}$ is the average temperature of the buffer layers; for the thermal mass storage in configurations 2 and 3 (section \ref{sec:goals_of_tests}), it represents an average indoor air temperature. This temperature is  constrained between a minimum and maximum limit  $\underline{T_{d}},\overline{T_{d}}$ (parameters).
These values are then used to construct a bid-function for every TCL \cite{ClaessensSelfLearning,Stijn,KlaasEventBased} expressed as the flow rate $L_{d}$ versus a heuristic ($p_{r}$). Above a corner value $p_{c,d}$ the bid function is zero:
\begin{equation}
p_{c,d}=1-SoC = \frac{\overline{T_{d}}-T_{d,t}}{\overline{T_{d}}-\underline{T_{d}}}.
\label{eq.3}
\end{equation}
Determining this heuristic is considered relatively straightforward as it requires only temperature measurements of the TLCs. As a first attempt, the flow rate ($L_d$) is estimated in the same way as described in section \ref{sec:reference_control}. Afterwards, the values of $L_d$ will be adapted through a PI-controller, as will be later discussed in Step 3. Finally the bid function for TCL $d$ has the following form: 
\begin{equation}
b_{d}(p_{r})=L_{d}(1-H(p_{r}+SoC-1)),
\label{eq.bidBoiler}
\end{equation}
where $H$ corresponds to the heaviside function. After determining the biding function for every TLC individually, these functions are aggregated (summarized).

A more detailed description can be found in \cite{Claessens2017}

\paragraph{Step 2: Optimization} ~\\
\label{s.optimization}
In the second  optimization step an optimal control vector $\mathbf{P}^{*}$ is determined, which is used as the control action in the third step. 
The objective of the energy arbitrage optimization $f(\mathbf{P})$ problem is defined as:
\begin{equation}
f(\mathbf{P}) = \sum_{t=1}^{T}{P_{t}\lambda_{t}\Delta t}+\sum_{t=1}^{T-1}{\alpha |P_{t+1}-P_{t}|}.
\label{eq.cost}
\end{equation} Here $\mathbf{\lambda}$ is the effective cost of supplying 1 unit of thermal energy, taking into account the cost for gas, reward obtained for producing electric energy, the fuel utilization ratio and the heat to power ratio \cite{Claessens2017}. $P_t$ is the total heating power delivered to cluster of TLCs, at time step $t$, which is the sum of power delivered to building thermal masses and the water tanks, as described below. The second term in eq. \ref{eq.cost} is added for regularization. 
The MPC models used in the optimization are simplified linear models, described below. \\
For the configurations where the thermal inertia of the buildings is used (the second and third configuration described in section \ref{sec:goals_of_tests}), an aggregated model is used to describe the dynamics of the entire cluster as in \cite{BiegelHP}. A second-order model \cite{Zhang} has been used as aggregated model. 
\begin{align}
\dot{T_{a}} & = \frac{1}{C_{a}} [T_{m}H_{m}-T_{a}(U_{a}+H_{m})+\gamma_{a}P_{b} Q_{a}+T_{out}U_{a}] \label{eq:Ta} \\
\dot{T_{m}} & = \frac{1}{C_{m}} [H_{m}(T_{a}-T_{m})+\gamma_{m}Q_{m}] \label{eq:Tm} 
\end{align}
Here $U_{a}$ is the conductance of the building envelope, $T_{out}$ the outside air temperature, $T_{a}= \sum_{d \in \mathcal{S}}{T_{d}}/|\mathcal{S}|$ the aggregated inside air temperature and $T_{m}$ the temperature of the thermal inertia. The conductance between $T_{a}$ and $T_{m}$ is represented by $H_{m}$. $Q_{a}$ and $Q_{m}$ represent the heat gains resulting from electric consumption and solar irradiance (these profiles are known inputs). $C_{a}$ and $C_{m}$ are the thermal mass of the air and the thermal inertia respectively. $P_{b}$ is the total power delivered to the cluster of buildings. Finally $\gamma_{a}$ and $\gamma_{m}$ represent scaling factors.
The model parameters are fitted based upon historic data, obtained by running the simulation for ten days. \\
When considering the water storage tanks configuration (the first and second configurations described in \ref{sec:goals_of_tests}) a first-order model has been used \cite{BiegelHP}.

\begin{equation}
\dot{T_{s}} = \frac{1}{C_{s}} [U_{s}(T_{out}-T_{s})+\gamma_{s}P_{w}-d]. \label{eq:Ts} \\
\end{equation} 

Here $C_{s}$ is the thermal capacity of the water, $T_{s}$ the average water temperature, $U_{s}$ the conductance of the storage tank, $d$ the off-take of thermal energy and $P_{w}$ is the heating power delivered to the water tanks. $\gamma_{s}$ a scaling factor. Also here these parameters are based upon historic data obtained by running the simulation for ten days. 

Finally the optimization performed every time step results in:
\begin{equation}
	\left.
	\begin{array}{l}
		\mathbf{P}^{*} ~= ~\underset{\mathbf{P}}{\mathrm{arg~min}} f(\mathbf{P}),\\
		\mathrm{s.t}.
		\left\{
			\begin{array}{l}
				\mathrm{eq.(}\ref{eq:Ta}\mathrm{)},\mathrm{eq.(}\ref{eq:Tm}\mathrm{)}\\      \mathrm{eq.(}\ref{eq:Ts}\mathrm{)}
			\end{array}
		\right.
	\end{array}
\right. 
\label{eq.secondOrder}
\end{equation}

This optimal power profile $\mathbf{P}^{*}$ is then used in the third step.

\begin{figure}[h!]
\centerline{\includegraphics[width=1.6\columnwidth]{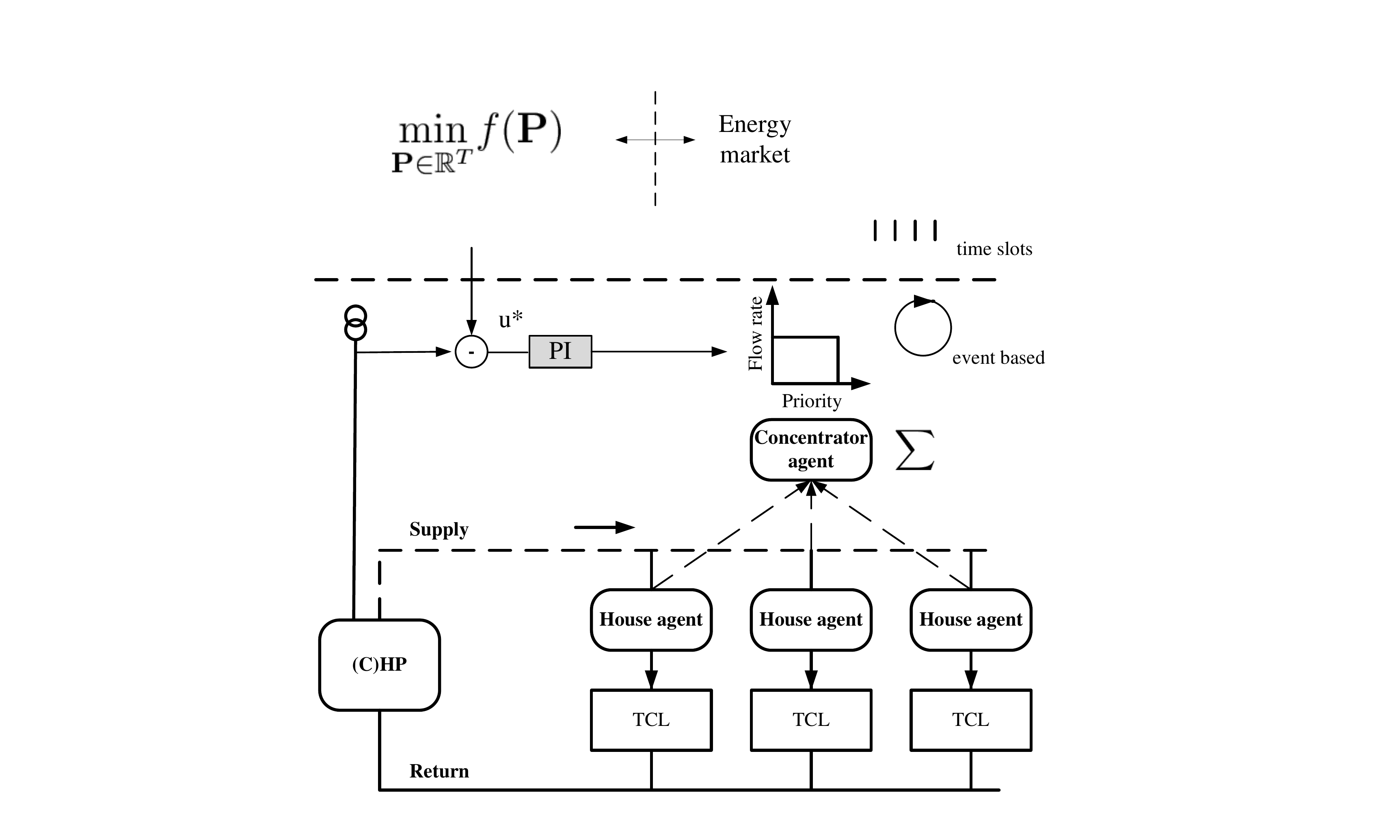}}
\caption{Overview of the controller approach as developed in this work.}
\label{fig.OverViewPV}
\end{figure}

\paragraph{Step 3: Real time control} ~\\
\label{s.realtime}
In the third step, the energy corresponding to $P_{t}^{*}$ is dispatched over the cluster of TCLs, using a market-based multi-agent system \cite{Stijn,Hommelberg,ClaessensSelfLearning}. As in \cite{BiegelHP}, a Proportional Integrator (PI) controller (at a central level) managing the flow rates at the different buildings, is used (Figure \ref{fig.OverViewPV}). This PI-controller will make sure that the actual consumed power by the cluster of TLCs is matched to the optimal power consumption defined by the MPC in Step 2.

The result of the PI controller $u$ is sent to the clearing process (\ref{eq.clearing}), after which a clearing priority ${p_{r}}^{*}$ is sent back to the different devices:
\begin{equation}
p_{r}^{*} ~= ~\underset{p_{r}}{\mathrm{arg~min}} \left|\sum_{d=1}^{|\mathcal{S}|}{b_{d}(p_{r}})-u\right|,
\label{eq.clearing}
\end{equation}
The devices open or close their local valve according to $b_{d}(p_{r}^{*})$.

\subsection{The tested week}
For every simulation, the same representative winter week was used. This week was chosen as the week in which the mean outside temperature was the closest to the mean temperature during the whole heating season in Belgium. A typical meteorological year (TMY) profile is used as weather data profile. The mean outside temperature during the heating season (1 October to 30 April) is $6.1^\circ C$. The week used in the simulation is the week with the smallest difference to this mean temperate. This is week 46 (12 to 18 November) with an mean ambient temperature of $6.2^\circ C$.

\section{RESULTS AND DISCUSSION}

\subsection{Operational behaviour of the district heating network components}
As discussed in the introduction, the aim of the active control algorithm is to maximise the revenue of the network operator. 

\begin{figure*}
	\centering
	\includegraphics[width = .9\textwidth]{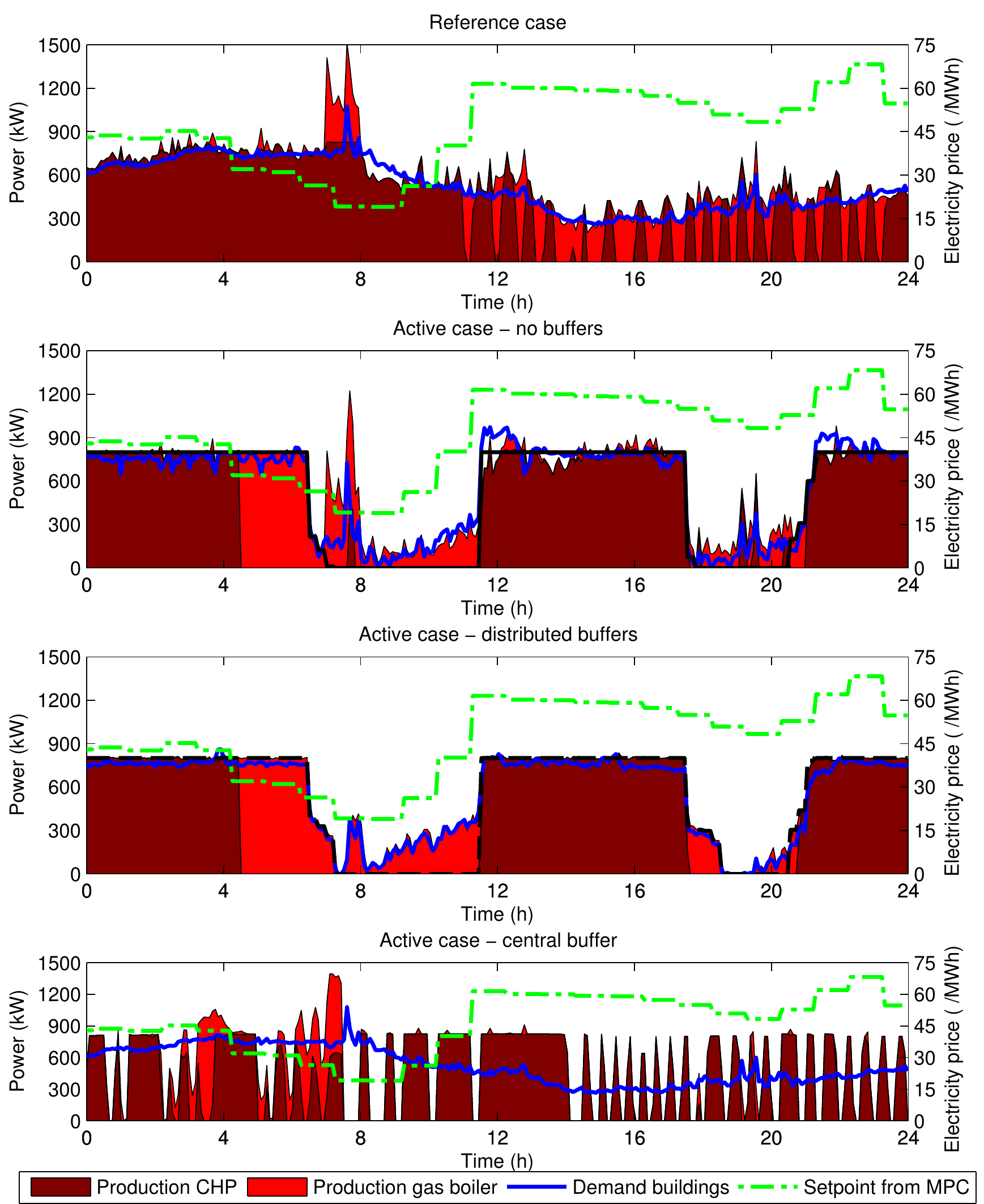}
	\caption{Power production and demand (left axis) and electricity price (right axis) during the tested week}
	\label{fig:powerflows}
\end{figure*}

Figure \ref{fig:powerflows} shows the behaviour in the DHN during one day (day 5). In the top figure the reference situation is shown. It is clear that the production and demand of heat (left axis) is independent of the price of electricity (right axis). There is a small deviation between heat production and heat demand because of heat losses in the grid and because of the time delay in the grid. If the power demand is higher than the lower modulation limit of the CHP, the CHP is switched on. If the power demand is below this limit, the CHP is switched off and the gas boiler takes over. This explains why after 11h the production is often fulfilled by the gas boiler. The graph also depicts (around 7h30) that, when the heat demand is higher than the maximum power of the CHP, the gas boiler supports the CHP.

The figures below show the same day for the active control cases. In these cases there is a much higher correlation between the heat production and high electricity prices (before 4h30, between 11h30 and 18h30 and after 20h30). As explained in \ref{sec:active_control_algorithm}, the controller framework consists of two different layers. Firstly, a MPC controller plans the optimal power production during the next time frame. The result of this planning is the black line in the figure.  Secondly, the MAS and PI controller distributes the thermal power to the most appropriate buildings. As can be seen, this controller performs properly, since the power production (the filled red area) corresponds very well to the planning. There are some deviations, e.g. for the active case without buffers between 7h30 and 11h30, where according to the planning the production should be 0. The reason for this is that when there is a domestic hot water demand, the district heating valve must be opened for comfort reasons, and therefore heat must be produced inevitably.

Once the optimal power production during the time step is known, the choice must be made of supplying the heat by the CHP or by the gas boiler. Since the CHP is selling its electricity to the spot market, the control algorithm will try to switch on the CHP when the electricity price is the highest. At other moments, when the electricity price is low, it could be more advantageous to switch on the gas boiler instead of the CHP. This can also be seen in the figure. In the active control case without buffers, before 5h, about 800 kW of heat must be produced. At that time the electricity price is high and therefore the demand is fulfilled by the CHP. However after 5h the electricity price decreases, and the heat is produced by the gas boiler. This phenomena can be observed a number of times.

\begin{figure}[h!]
	\centering
	\includegraphics[width=.4\textwidth]{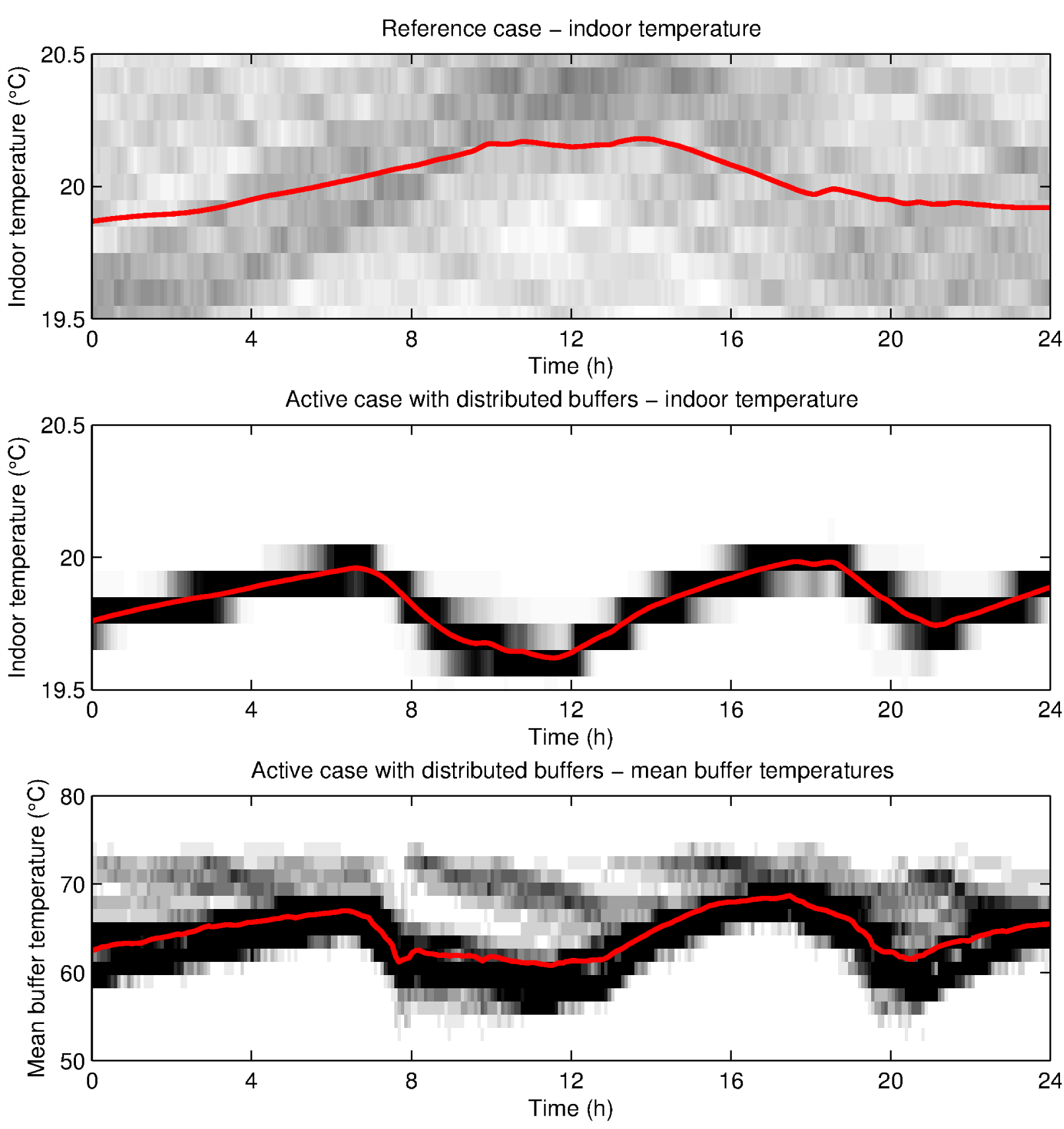}
	\caption{Building and buffer behaviour for the reference and active control case with 	distributed buffers}
	\label{fig:buildingbehaviour}
\end{figure}

Figure \ref{fig:buildingbehaviour} gives an insight on how the heat demand profile is manipulated by the control framework, for the active case with distributed buffers. As explained above, in this case also the thermal mass of the buildings is used to store heat. The top figure shows the mean indoor temperature of all the buildings (red line) and the distribution of this indoor temperature (shading) for the reference case. Comparing the indoor temperature of the reference case to this of the active case (figure below), one can notice that the indoor temperature distribution for the active case is much more concentrated. This is because the MAS controller always distributes the heat available to the house with the highest heat demand. As a consequence, the building with the lowest temperature is always dragged to the rest of the buildings. In the last figure, the mean temperature in the distributed buffers is shown. A comparable evolution as for the indoor temperature can be noticed.

\subsection{Energy consumption and production}
\label{sec:total_energy}
In table \ref{tab:energy} the total energy consumption and production is shown for the different configurations. 

\setlength{\tabcolsep}{0pt}
\begin{table*}[h!]
\begin{center}
\begin{tabular}{p{6cm} p{1.62cm} p{1.62cm} p{1.62cm} p{1.62cm} p{1.62cm} p{1.62cm} p{1.62cm}  p{1.62cm}}
\hline
 & \multicolumn{2}{c}{energy consumed} & \multicolumn{2}{c}{energy produced} & \multicolumn{2}{c}{energy produced} & \multicolumn{2}{c}{energy produced} \\
configuration & \multicolumn{2}{c}{} & \multicolumn{2}{c}{total} & \multicolumn{2}{c}{CHP} & \multicolumn{2}{c}{gas boiler}	\\
\hline
no buffers, regular control (reference) & 70649 & & 79447 & & 58714 & & 20651 & \\
no buffers, active control & 73562 & $(+4.1\%)$ & 79600 & $(+1.9\%)$ & 52100 & $(-11.3\%)$ & 27985 & $(+35.5\%)$ \\
distributed buffers, active control & 73594 & $(+4.2\%)$ & 80965 & $(-0.2\%)$ & 59177 & $(+0.8\%)$ & 21900 & $(+6.0\%)$ \\
central buffer, active control & 70577 & $(-0.1\%)$ & 78741 & $(-0.9\%)$ & 43750 & $(-25.6\%)$ & 35804 & $(+73.4\%)$ \\
\hline 
\end{tabular}
\caption{Produced and consumed energy for the different configurations (in kWh) and difference in terms of percentage to the reference case }
\label{tab:energy}
\end{center}
\end{table*}
\setlength{\tabcolsep}{8pt}

With respect to the energy demand of the buildings, one can see that the consumption of the central buffer is the same as for the reference case. This is logical, since in both cases the control strategy of the individual buildings is the same. For the other active cases, the energy demand of the buildings is a higher than that of the reference control case. The difference is the result of the different control strategy. In the reference case, the number of times that the district heating valve is opened, is minimized: only when the indoor temperature drops below the lower limit ($19.5^\circ C$) the valve is opened until the indoor temperature reaches the upper limit ($20.5^\circ C$). This valve is opened a lot more in the active cases. For the configuration with the distributed buffers the higher consumption is also resulting from the heat loss of the distributed buffers. The total heat loss of these buffers amounts to $606 kWh$ during the tested week.

\begin{figure*}
	\centering
	\includegraphics[width = .9\textwidth]{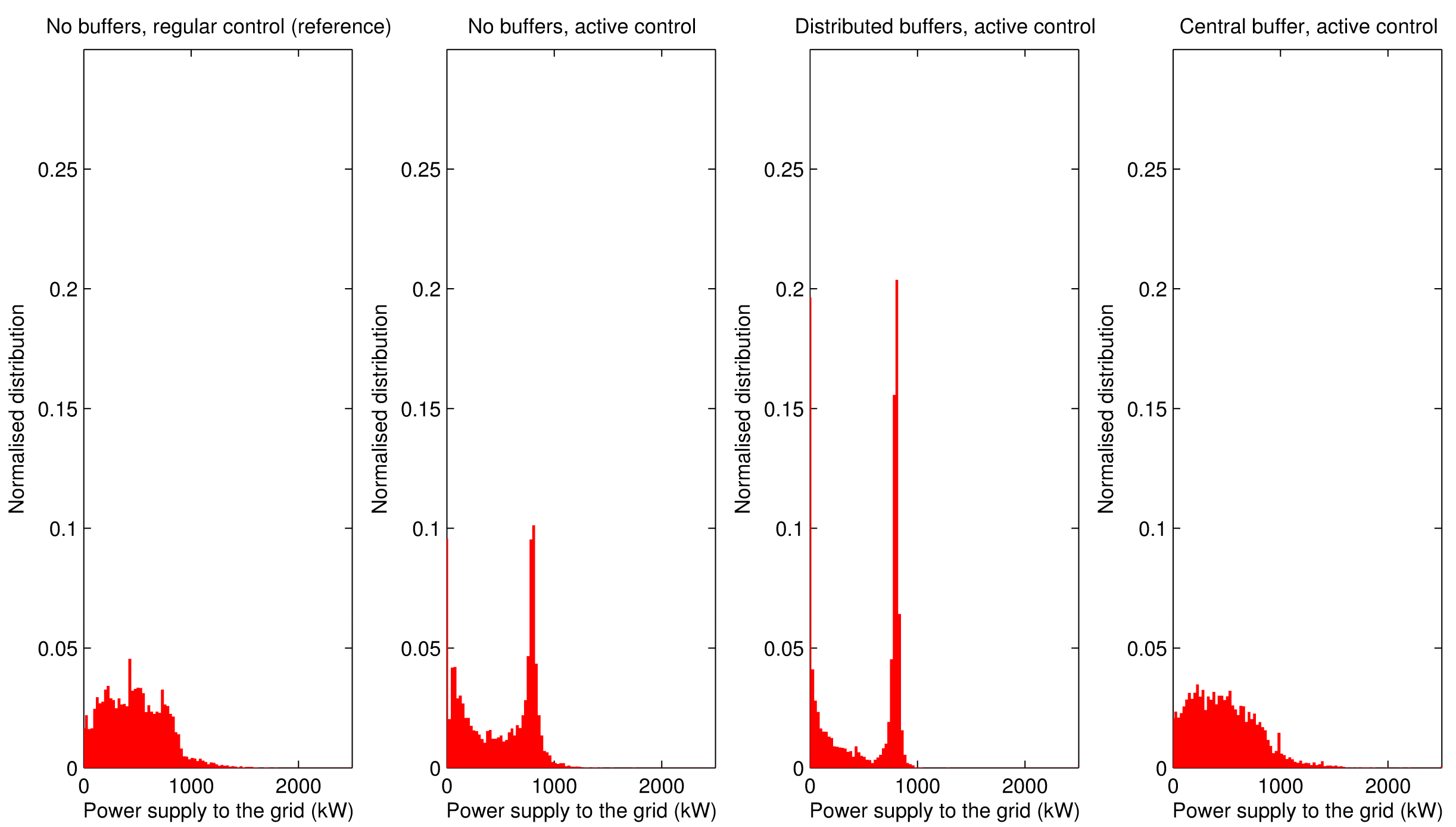}
	\caption{Distribution of the power supply to the grid for the different storage configurations}
	\label{fig:hist_power}
\end{figure*}

Having a look at the production of energy, i.e. the energy which is delivered to the district heating grid, compared to the reference case, the efficiency of the district heating grid is somewhat higher for the active control case with distributed buffers and the active case without buffers ($90.9\%$ and $92.4\%$ versus $88.9\%$). This means that the heat losses in these cases would be a bit lower. This is explained by the variation of the power demand and power supply in the grid. In the reference case, whenever one of the buildings need heat, the district heating valve is opened. This means that often the total heat demand of the grid is low. In the active control cases however, the algorithm aims to switch off the power demand of the buildings until the electricity price is high. Once the price is high enough the CHP should switch on, implying that the energy demand of the buildings should be higher than the lower modulation limit of the CHP. Therefore, less moderate heat demands occur in the active cases: the heat demand is either high or very low, as can be seen in the histograms in figure \ref{fig:hist_power}.

Higher powers correspond to higher flow rates and velocities, and shorter travelling times for the water through the pipes, and consequently lower heat losses. As can be seen from the figure, the power distribution of the active case with a central buffer is comparable to the distribution of the reference case. Therefore also the efficiency of this case is about the same ($89.6\%$).

\subsection{Costs and revenues}
In this final analysis the costs and revenues of the different configurations are compared. The costs consist of the gas costs of the CHP and the gas boiler and the pumping costs. The revenues are resulting from the sale of heat to the customers and electricity to the spot market. The parameters used in this analysis are: a natural gas price of \euro $39.9/MWh$ (price of gas for small industrial consumers in Belgium) and a price for heat sold of \euro $54.5/MWh$ (price of gas for residential consumers in Belgium). The price profile of electricity sold is the same profile as the one shown in Figure \ref{fig:buildingbehaviour} and represents the Belgian wholesale price.

\begin{figure}[h!]
	\centering
	\includegraphics[width = .4\textwidth]{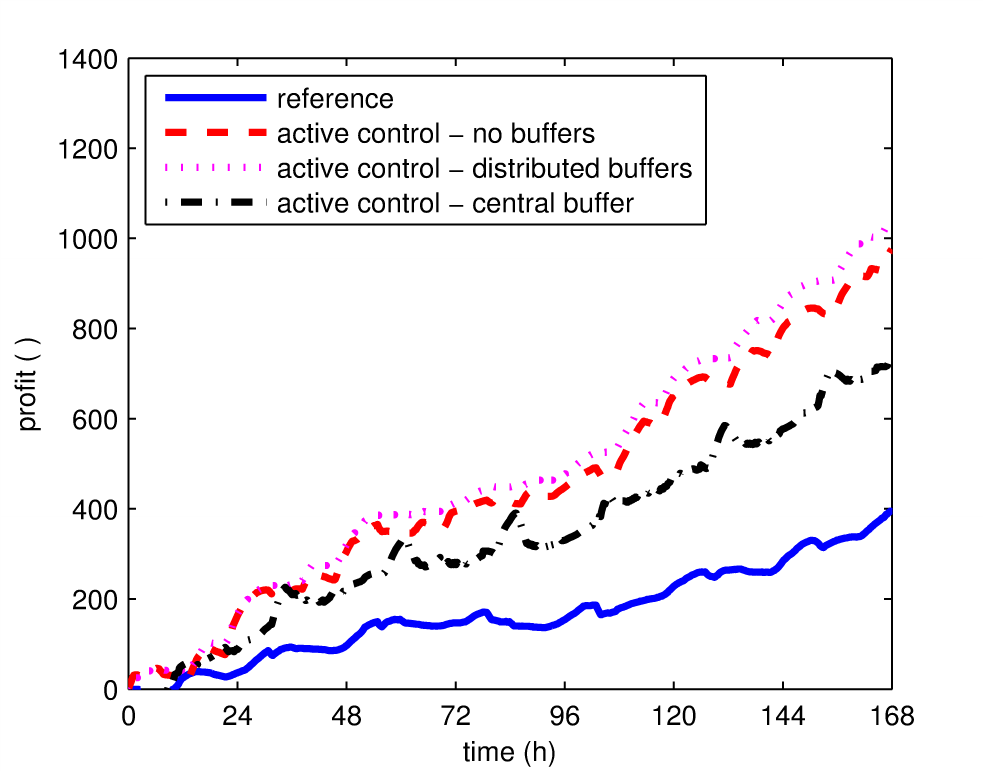}
	\caption{Operational profit for the different buffer configurations during the tested week}
	\label{fig:profit}
\end{figure}

In figure \ref{fig:profit} the operational profit - here defined as the difference of the total revenue minus the total costs - is shown for the different configurations. As can be seen, the active control is able to significantly increase this profit due to the higher revenue from the electricity production of the CHP.

This analysis also indicates that the active case with distributed buffers performs best. The case with distributed buffers performs slightly worse. The case with the central buffer gives less good results, however the difference with the reference case still is significant. This is a bit surprising, since one would expect higher flexibility for the configuration with a central buffer, as also stated in the Nuytten \cite{Nuytten2013}. This cited work nevertheless only calculated the flexibility of the system, not how well that flexibility is utilised. Moreover, a large difference between the central buffer configuration and the other active configurations is that in these last configurations also the building mass of the houses is activated. The total thermal mass of the buildings in the simulation amounts to $4165 kWh/K$, which is a lot higher than the thermal mass of the central buffer and the distributed buffers (both $39.5 kWh/K$). Raising the building indoor temperature by $1^\circ C$ therefore corresponds a temperature rise in the buffers of more than $100^\circ C$. Activating the building mass is therefore shown to be very interesting. For the same reason, the difference in profit between the active case without buffers and the case with distributed buffers is limited.

\section{CONCLUSIONS}
In this work a number of storage configurations for district heating grids are compared to each other. To achieve this, a simulation model was built. The flexibility resulting from the storage vessels is used to actively control a CHP, which in this way can produce electricity at times of high electricity prices. 
The simulation results indicate that the developed control framework perform well, i.e. that the business case – maximisation of the profit – is achieved.
The profit resulting from the different storage configurations are compared to each other. The results show that active control of the CHP is able to increase the profit of the CHP significantly. The configuration with distributed buffers performs best, however only slightly better than the active configuration without buffers. The results for the central buffer case are a little worse, but still a lot better than in the reference case. The reasons for this worse behaviour is that the thermal mass of the buildings, which is activated by the first two configurations, is a lot higher than that of the buffers, resulting in much more flexibility and as a result higher yields.

\section*{Acknowledgements}
This work has been done under the FP-7 program E-Hub (260165) and FC-District (260105).

\section*{References}
\bibliography{distributed_storage_paper}

\end{document}